\documentclass[11pt,a4paper]{article}
\usepackage{amssymb,amsmath,latexsym,enumerate,graphicx,microtype,cite}
\usepackage{abstract}
\allowdisplaybreaks
\usepackage{a4}

\usepackage{xspace}
\usepackage{amsmath,amssymb,amsthm,mathrsfs}
\usepackage{mathtools}
\usepackage{comment}
\usepackage{abstract}
\usepackage{subfigure}

\usepackage[colorinlistoftodos]{todonotes}

\renewcommand{\leq}{\leqslant}
\renewcommand{\geq}{\geqslant}

\def\zplus{{\mathbb{Z}_{\geq 0}^d}}

\newcommand{\DSP}{\emph{DSP}\xspace}

\def\aov{{\overrightarrow{\bf{1}}}}

\newcommand{\Zon}{\mathscr{Z}}

\newtheorem{theorem}{Theorem}

\newtheorem{lemma}{Lemma}
\newtheorem{proposition}{Proposition}
\newtheorem{observation}{Observation}
\newtheorem{claim}{Claim}

\date{May 11, 2023}

\author{Eleonore Bach \thanks{EPFL, Lausanne, Switzerland} , Friedrich Eisenbrand \thanks{EPFL, Lausanne, Switzerland} , Rom Pinchasi \thanks{EPFL, Lausanne, Switzerland and Technion, Haifa, Israel}}

\title{Integer points in the degree-sequence polytope}

\begin{document}
\maketitle

\begin{abstract}
An integer vector $b \in \mathbb{Z}^d$ is a \emph{degree sequence} if there exists a hypergraph with vertices $\{1,\dots,d\}$ such that each  $b_i$ is the number of hyperedges containing~$i$. 
The \emph{degree-sequence polytope} $\Zon^d$  is the convex hull of all degree sequences.
 We show that all but a $2^{-\Omega(d)}$ fraction of integer vectors in the degree sequence polytope are degree sequences.  Furthermore, the corresponding hypergraph of these points can be computed in time $2^{O(d)}$ via linear programming techniques. This is substantially faster than the  $2^{O(d^2)}$ running time of the current-best algorithm for the degree-sequence  problem.  
We also show that for  $d\geq 98$, $\Zon^d$ contains integer points that are not degree sequences.
Furthermore, we prove that the linear optimization problem over $\Zon^d$ is $\mathrm{NP}$-hard. This complements a recent result of Deza et al. (2018) who provide an algorithm that is polynomial in $d$ and the number of hyperedges. 
\end{abstract}

\section{Introduction}
\label{sec:introduction}

A \emph{hypergraph}~\cite{berge1984hypergraphs,bollobas1986combinatorics}  is a tuple $H = (V,E)$   where $V = \{1,\dots,d\}$ is the set of \emph{vertices} and $E \subseteq 2^V$ is the set of hyperedges. The \emph{degree} of a vertex $v \in V$ is the number of hyperedges containing $v$ and the  \emph{degree sequence} of $H$ is the vector $b = (b_1,\dots,b_d)^{\top} \in \mathbb{N}^d$ where $b_i$ is the degree of vertex $i$.
This paper focuses on the complexity of the following decision problem to which we refer to as the \emph{degree-sequence problem} (\DSP). 
\begin{quote}
  \label{prob:1}
  Given a vector $b \in\mathbb{N}^d$, is it a degree sequence of some hypergraph on $d$ vertices? 
\end{quote}

The fastest known algorithm for \DSP is the following dynamic program that runs  in time $2^{O(d^2)}$. Let $\chi_1,\dots,\chi_{2^d-1} \in  \{0,1\}^d$  be the characteristic vectors of the subsets of $\{1,\dots d \}$, ordered in an arbitrary way and consider a directed  acyclic graph with vertices
\begin{align*}
  \left\{ (i,x) : 0 \leq  i \leq 2^d-1, \,x  \in  \mathbb{N}_0^d, \ \|x\|_{\infty} \leq 2^{d-1} \right\}.      
\end{align*}
There is an arc between each ordered pair of vertices  $(i,x)$ and  $(i+1,x)$
as well as between  $(i,x)$ to $(i+1,x + \chi_{i+1})$. These arcs represent not-choosing/choosing  the $i+1$-st hyperedge. A vector $b$  is a degree sequence if and only if there exists a path from $(0,0)$ to $(2^d-1,b)$ in this graph. 

\subsubsection*{Integer programming}
This running time matches  the state-of-the-art~\cite{papadimitriou1981complexity,EisWeis} for 
 \emph{integer programming} 
\begin{equation}
  \label{eq:3}
  \max \{ c^{\top} x : A\, x = b,\, x \in \{0,1\}^n \} 
\end{equation}
with $0/1$-constraint matrix $A \in \{0,1\}^{d \times n }$.  Notice that \DSP is the feasibility problem for~\eqref{eq:3} with the \emph{complete} $0/1$-constraint matrix $A$ whose columns are all possible $0/1$-vectors.  

The version of~\eqref{eq:3} without upper bounds on the variables, i.e., $x  \in  \mathbb{Z}_{\geq 0}^n$ can be solved in time $2^{O(d)}$~\cite{EisWeis,jansen2022integer}. 
Knop et al.~\cite{knop2020tight} have provided a matching lower bound of $2^{\Omega(d)}$ on the running time of any algorithm solving an integer program~\eqref{eq:3}. The lower bound is based on the \emph{Exponential Time Hypothesis}~\cite{impagliazzo2001complexity} which has been established as a tool to provide lower bounds for many algorithmic problems, see, e.g.~\cite{cygan2015parameterized}.

Whether an integer program~\eqref{eq:3} with $0/1$ constraint matrix  can be solved in \emph{single exponential time}, i.e., in time $2^{O(d)}$, is a prominent open problem. A positive resolution would have immediate impact for example  in the area of fixed-parameter tractability and scheduling~\cite{jansen2022empowering,jansen2020approximation,knop2018scheduling}, which is currently vibrant.

\subsection*{Contributions of this paper} 

The \emph{degree-sequence polytope}~\cite{WhiteWhale,Stanley,MS} is
the convex hull $\Zon^d$ of the degree sequences of hypergraphs with vertex set $\{1,\dots,d\}$.  This corresponds to the $d$-dimensional zonotope generated by the $2^d$ distinct $\{0,1\}^d$ vectors
\begin{equation}
  \label{eq:1}
  \Zon^d = \{ A^{(d)} x : x  \in  [0,1]^{2^d-1} \}
\end{equation}
where $A^{(d)}  \in  \{0,1\}^{d \times 2^d-1}$  is the \emph{complete binary matrix}, whose columns consist of all possible nonzero $0/1$-vectors.

{\bf 
Our main result} shows that all but an
  \emph{exponentially small} fraction of integer vectors in the degree-sequence polytope $\Zon^d$
  are degree sequences. More precisely, candidates  $b  \in  \Zon^d \cap \mathbb{Z}^d$  that are not degree sequences lie very close to the boundary.
 This result implies that an integer point $b$ chosen i.i.d. at random from $ \Zon^d \cap \mathbb{Z}^d$ is a degree sequence with a very high probability of at least
 \begin{equation}
   \label{eq:2}   
   1-\theta\biggl(\frac{d^2}{2^{d-1}}\biggl).
 \end{equation}

This has the following algorithmic consequence. For a random vector $b  \in  \mathbb{N}_0^d$ with   $\|b\|_{\infty} \leq 2^{d-1}$, with probability~\eqref{eq:2}, we can decide whether $b$ is a degree sequence in time $2^{O(d)}$ by solving the linear program $b  \in  \Zon^d$~\cite{khachiyan1979polynomial}. 

This is  substantially faster than the  $2^{O(d^2)}$ dynamic program. 
   
We show for $d \geq 98$ that there are integer points in $\Zon^d$ that are not degree sequences (Theorem \ref{theorem:main}). For  $d \leq 9$ computer experiments show that \emph{all} integer
points in $\Zon^d$ are degree sequences.

Finally, we  provide the following complexity results:
\begin{itemize}
\item[-] 
The degree sequence problem is 
$\mathrm{NP}$-hard. This is via a reduction from the result of Deza et
al.~\cite{MR3840885} who show that it is hard to decide the degree
sequence problem for $3$-uniform hypergraphs.  A hypergraph is
\emph{$k$-uniform} if each of its hyperedges contains $k$ elements.
\item[-]
   The linear optimization problem over
   $\Zon^d$ is $\mathrm{NP}$-hard, as well. This is contrasting a
   result of Deza et al.~\cite{MR3840885} who showed that
   the linear optimization problem over degree sequences can be solved
   in polynomial time in $d$ and the number $m$ of hyperedges.
 \end{itemize}

\subsection*{Related results}

The degree sequence problem has received considerable attention in the
literature. 
A classical result of
Erd\H{o}s and Gallai \cite{erdosgallai} characterizes the degree
sequences of $2$-uniform hypergraphs, i.e., standard undirected
graphs. This characterization permits to decide the
 degree-sequence problem for $2$-uniform graphs in polynomial time. 
The problem whether a similar characterization exists for $3$-uniform
hypergraphs was open for more than 30 years~\cite{Colbourn} until Deza et
al.~\cite{MR3840885,deza2019hypergraphic} showed that, for each $k\geq 3$ fixed,  the degree-sequence problem for
$k$-uniform hypergraphs is $\mathrm{NP}$-complete.  \footnote{It is worth noting  that it is not clear whether the degree-sequence problem~\eqref{prob:1}  is in
  $\mathrm{NP}$ or $\mathrm{co-NP}$ if the encoding of the vector
  $b \in \mathbb{N}^d$ is in binary.} 
Liu \cite{Liu} showed that the set of $k$-graphic sequences is non-convex, implying that Erd\H{o}s-Gallai-type theorems do not hold. Hence, a main focus was to find good Havel-Hakimi \cite{Havel, Hakimi} like generalizations to $k$-uniform hypergraphs.

\section{Almost all integer points in the degree sequence-polytope are degree sequences}

We denote by $C_{d}$ the set of vertices of the discrete $d$-dimensional cube and we denote by $A$ the complete binary matrix $A$ of size $d$. This is the $d \times 2^d$ matrix whose columns 
are the vectors in $C_{d}$.
Note that we can write $\Zon^d$ as the Minskowski sum 
of the $2^d$ straight line segments $[0, v]$, where $v \in C_d$ 
is any of the $2^d$ vertices of the discrete $d$-dimensional cube.
From this representation we directly see that $\Zon^d$ is indeed a zonotope. 

Let us denote by $S$ the set of degree sequences points of the
degree-sequence polytope $\Zon^d$. 

As we will see in section \ref{sec:non-realizable}, $\Zon^d$ contains more
integer points than those that are in $S$.
In this section we show that most of the integer points of the degree-sequence polytope are already points in $S$.
More precisely, we show below in Theorem \ref{theorem:boundary} that all integer points in $\Zon^d$
whose distance from the boundary of $\Zon^d$ 
is greater than $\sqrt{d}(d+1)^2$ are degree-sequences.  We recall that the diameter of $\Zon^d$
is $2^{d-1}\sqrt{d}$. The smallest distance of a boundary point of $\Zon^d$ to the center
of $\Zon^d$ is $\theta(2^{d-1}\sqrt{d})$. Hence, by showing that all integer points in $\Zon^d$
whose distance from the boundary of the convex hull of $\Zon^d$ 
is greater than $\sqrt{d}(d+1)^2$ are degree sequences, we may conclude that 
the ratio of degree sequences  to all the
integer points in the degree-sequence polytope is $\theta((1-\frac{(d+1)^2}{2^{d-1}})^d)=
1-\theta(\frac{d^2}{2^{d-1}})$.

 Let $e_i$, $1 \leq i \leq d$, denote the $i$-th unit vector in $\mathbb{R}^d$. Let $\aov$ denote the all-ones vector in $\mathbb{R}^d$. That is, $\aov=e_{1}+ \ldots+e_{d}$. 
 We will need the following simple observation about $S$ and $\Zon_{d}$.

\begin{observation}\label{observation:symmetric}
Both $S$ and $\Zon_{d}$ are centrally symmetric about the point
$2^{d-2}\aov$. 
\end{observation}

\begin{proof}

Indeed, assume that $P \in \Zon_{d}$ (resp., $P \in S$). Therefore, 
$P=\sum_{v \in C_{d}}a_{v}v$, where $0 \leq a_{v} \leq 1$ 
(resp., $a_{v} \in \{0,1\}$).
Consider the point $Q=\sum_{v \in C_{d}}(1-a_{v})v$. Clearly, $Q \in \Zon_{d}$ (resp. $Q \in S$). 
We have
$Q=(\sum_{v \in C_{d}}v)-P=2^{d-1}\aov-P=2(2^{d-2}\aov)-P$, showing that 
$2^{d-2}\aov$ is the midpoint of the straight line segment determined by
$P$ and $Q$. 
\end{proof}

Our main result in this section can be summarized in the following theorem.

\begin{theorem}\label{theorem:boundary}
Let $b \in \Zon^d$ be an integer point whose distance from the boundary of 
$\Zon^d$ is greater than $\sqrt{d}(d+1)^2$. Then $b \in S$.
\end{theorem}

Theorem \ref{theorem:boundary} will be a direct consequence of 
Theorem \ref{theorem:F}, Theorem \ref{theorem:F2}, Theorem 
\ref{theorem:main3}, and Theorem \ref{theorem:main2} below.

\begin{theorem}\label{theorem:F}
Assume $b=(b_{1}, \ldots, b_{d})$ is an integer point such that $b-d^2\aov \in 
\Zon^d$. Assume also that $b_{i} \leq 2^{d-2}$ for every $1 \leq i \leq d$.
Then $b \in S$.
\end{theorem}

As an immediate consequence of Theorem \ref{theorem:F} we get the
following symmetric theorem for the case $b_{i} \geq 2^{d-2}$ for every 
$1 \leq i \leq d$.

\begin{theorem}\label{theorem:F2}
Assume $b=(b_{1}, \ldots, b_{d})$ is an integer point such that $b+d^2\aov \in 
\Zon^d$. Assume also that $b_{i} \geq 2^{d-2}$ for every $1 \leq i \leq d$.
Then $b \in S$.
\end{theorem}

\begin{proof} By Observation \ref{observation:symmetric} and 
the assumption that $b+d^2\aov \in \Zon^d$, we have
$(2^{d-1}\aov -b) - d^2\aov=2^{d-1}\aov-(b+d^2\aov) \in \Zon^d$. 
We apply Theorem \ref{theorem:F} for the vector $2^{d-1}-b$ to conclude that
$2^{d-1}\aov-b \in S$. Again, by Observation \ref{observation:symmetric}, $b = 2^{d-1}\aov-(2^{d-1}\aov-b) \in S$, as desired.
\end{proof}

With the following theorems we extend Theorem \ref{theorem:F} and Theorem 
\ref{theorem:F2} to the more general
case where some of the coordinates of $b$ may be greater than $2^{d-2}$ and some are smaller than $2^{d-2}$.

\begin{theorem}\label{theorem:main3}
Assume $b=(b_{1}, \ldots, b_{d})$ is an integer point and $I \cup J$
is a partitioning of $\{1, \ldots, d\}$ with $|I| \geq |J|>0$ such that 
$b_{i} \leq 2^{d-2}$ for every $i \in I$ and $b_{j} > 2^{d-2}$
for every $j \in J$. 
If $b-(d+1)^2(\sum_{i \in I}e_{i}-\sum_{j\in J}e_{j})) \in 
\Zon^d$, then $b \in S$.
\end{theorem}

As a direct consequence of Theorem \ref{theorem:main3} we get 
the following similar statement in the case $0<|I| < |J|$.

\begin{theorem}\label{theorem:main2}
Assume $b=(b_{1}, \ldots, b_{d})$ is an integer point and $I \cup J$
is a partitioning of $\{1, \ldots, d\}$ with $0<|I| < |J|$ such that 
$b_{i} < 2^{d-2}$ for every $i \in I$ and $b_{j} \geq 2^{d-2}$
for every $j \in J$. 
If $b-(d+1)^2(\sum_{i \in I}e_{i}-\sum_{j\in J}e_{j})) \in 
\Zon^d$, then $b \in S$.
\end{theorem} 

\begin{proof}   
Consider the vector
$b'=2^{d-1}\aov-b$. Writing $b'=(b'_{1}, \ldots, b'_{d})$ we show that $I$ and $J$
interchange roles and we have $b'_{i} > 2^{d-2}$ for every $i \in I$ and 
$b'_{j} < 2^{d-2}$ for every $j \in J$. 

If we assume that $b-(d+1)^2(\sum_{i \in I}e_{i}-\sum_{j\in J}e_{j})
\in \Zon^d$, then, by Observation \ref{observation:symmetric}, we have
$$
b'-(d+1)^2(\sum_{j\in J}e_{j}-\sum_{i \in I}e_{i})=
2^{d-1}\aov-(b-(d+1)^2(\sum_{i \in I}e_{i}-\sum_{j\in J}e_{j})) \in \Zon^d.
$$

We can now use Theorem \ref{theorem:main3} and conclude that $b' \in S$.
Now we apply Observation \ref{observation:symmetric} once again to obtain
$b=2^{d-1}\aov-b'\in S$, as desired.
\end{proof} 

\subsection{Some Preliminary Tools}

We start by developing some tools needed for the proofs of 
Theorem \ref{theorem:F} and Theorem \ref{theorem:main3}.

\begin{claim}\label{claim:1}
Suppose $v=(v_{1}, \ldots, v_{d}) \in S$ and $v_{i} < 2^{d-2}$ for 
some $1 \leq i \leq d$. Then we have 
$v+\aov+e_{i} \in S$.
\end{claim}

\begin{proof}
    
There are $2^{d-1}$ different ways to write $\aov+e_{i}$ as a sum $f+g$
of two distinct vectors $f,g \in C_{d}$. Because such $f$ and $g$ determine each
other in a unique way ($f+g=\aov+e_{i}$), we get a partitioning of 
the $2^{d-1}$ vectors in $C_{d}$, whose $i$-th coordinate is equal to one,
into $2^{d-2}$ unordered pairs of the form $\{f,g\}$, where $f+g=\aov+e_{i}$.
We call such an unordered pair $\{f,g\}$ a \emph{good} pair.

Because $v \in S$, we can write $v=u_{1}+\ldots+u_{k}$ where $u_{1}, \ldots, u_{k}$
are distinct vectors in $C_{d}$. Since $v_{i} < 2^{d-2}$, 
there must be a good pair $\{f,g\}$ such that neither $f$ nor $g$, is one
of $u_{1}, \ldots, u_{k}$. Hence we can write 
$v+\aov+e_{i}=u_{1}+\ldots +u_{k}+f+g$, showing that $v+\aov+e_{i}$ is in $S$.
\end{proof} 

\begin{claim}\label{claim:2}
Suppose $v=(v_{1}, \ldots, v_{d}) \in S$ and $v_{i} < 2^{d-2}$ for 
every $1 \leq i \leq d$. Then we have 
$v+\aov \in S$.
\end{claim}

\begin{proof}

Write $v=u_{1}+ \ldots +u_{k}$, where $u_{1}, \ldots, u_{k}$ are pairwise distinct
vectors in $C_{d}$. 

If there are $1 \leq i \leq d$ and $1 \leq j \leq k$ such that 
the $i$-th coordinate of $u_{j}$ is equal to $1$, but $u_{j}-e_{i}$ is not 
one of the vectors in the sum representing $v$, then we are done.
Indeed, replace $u_{j}$ with $u_{j}-e_{i}$ and observe that $v-e_{i} \in S$.
Then apply Claim \ref{claim:1}
to obtain that $v+\aov=v-e_{i}+(\aov+e_{i}) \in S$.

We claim that there must be such $i$ and $j$ by considering the vector $u_{j}$
with the least number of coordinates that are equal to one. Then every
$i$ such that the $i$-th coordinate of $u_{j}$ is equal to one is good.
This concludes the proof.
\end{proof}

\begin{observation}\label{observation:simple}
Assume $q=(q_{1}, \ldots, q_{d}) \in S$ and $q_{i}>q_{j}$ for some 
$i$ and $j$. Let $q'=(q'_{1}, \ldots, q'_{d})$ be defined as follows:  
$q'_{\ell}=q_{\ell}$ for every $\ell \neq i,j$ while $q'_{i}=q_{i}-1$ and
$q'_{j}=q_{j}+1$. Then $q' \in S$.
\end{observation}

\begin{proof}
Because $q \in S$ we can write $q=u_{1}+\ldots+u_{k}$ where $u_{1}, \ldots, u_{k}$
are pairwise distinct vectors in $C_{d}$. 
Because $q_{i}>q_{j}$ there must be $u_{\ell}$ such that 
the $i$-th coordinate of $u_{\ell}$ is equal to one, the $j$-th coordinate
of $u_{\ell}$ is equal to zero, and $u_{\ell}-e_{i}+e_{j}$ is not one of 
$u_{1}, \ldots, u_{k}$. Otherwise the number of vectors
among $u_{1}, \ldots, u_{k}$ in which the $i$-th coordinate is greater than
the $j$-th coordinate would not be greater than the number of vectors
among $u_{1}, \ldots, u_{k}$ in which the $i$-th coordinate is smaller than
the $j$-th coordinate, contradicting the assumption that $q_{i}>q_{j}$.

Now we observe that
$u_{1}, \ldots, u_{k}$ with $u_{\ell}$ replaced with $u_{\ell}-e_{i}+e_{j}$ is
a collection of pairwise distinct vectors in $C_{d}$ whose sum is equal to $q'$.
Hence $q' \in S$.
\end{proof}

\begin{lemma}\label{lemma:simple}
Assume $q=(q_{1}, \ldots, q_{d})\in S$, $b=(b_{1}, \ldots, b_{d})$ is an 
integer vector and there exists a partition $I \cup J$ of 
$\{1, \ldots, d\}$ such that the following three conditions are satisfied

\begin{itemize}
\item[(1)]\label{c1} $b_{i} \leq b_{j}$ for every $i \in I$ and $j \in J$.

\item[(2)]\label{c2} $q_{i} \leq b_{i}$ for every $i \in I$
and $q_{j} \geq b_{j}$ for every $j \in J$.

\item[(3)]\label{c3} $\sum_{i=1}^{d}b_{i}=\sum_{i=1}^{d}q_{i}$. 
\end{itemize}

Then $b \in S$.
\end{lemma}

\begin{proof}
    
We prove the lemma by induction on 
$\sum_{j \in J}(q_{j}-b_{j})$.

If $\sum_{j \in J}(q_{j}-b_{j})=0$, then also 
$\sum_{i \in I}(q_{i}-b_{i})=0$ because of the Conditions (1) and (3).
Because of Condition (2) we conclude that $q_{i}-b_{i}=0$ for every 
$1 \leq i \leq d$ and hence $b=q \in S$.

Assume now that $\sum_{j \in J}(q_{j}-b_{j})>0$.
Then there is $j \in J$ such that $q_{j}>b_{j}$.
Because $\sum_{i=1}^{d}b_{i}=\sum_{i=1}^{d}q_{i}$, the
sum $\sum_{i \in I}(q_{i}-b_{i})<0$. Consequently, there exists an
$i \in I$ such that $q_{i}<b_{i}$. 
By Observation \ref{observation:simple}, $q-e_{j}+e_{i} \in S$.
We notice that $q-e_{j}+e_{i}$ and $b$ satisfy the conditions of Lemma 
\ref{lemma:simple}. We can now use the induction hypothesis and conclude that 
$b \in S$. 
\end{proof}

\begin{claim}\label{claim:modulu}
Assume $P=(p_{1}, \ldots, p_{d}) \in S$ and $\sum_{i=1}^{d}p_{i}>d+1$.
Let $0 \leq r<d+1$ be an integer.
Then there exists $Q=(q_{1}, \ldots, q_{d})$ such that 
$q_{i} \leq p_{i}$ for every $1 \leq i \leq d$, 
$\sum_{i=1}^{d}p_{i}=r+\sum_{i=1}^{d}q_{i}$ and $Q \in S$.
\end{claim}

\begin{proof}
Because $P \in S$ we can write $P=u_{1}+ \ldots +u_{k}$, where
$u_{1}, \ldots, u_{k}$ are pairwise distinct vectors in $C_{d}$.

We prove the theorem by induction on $r$.
For $r=0$ we can take $Q=P$. Assume $r>0$ and consider $u_{\ell}$ such that the number of coordinates
of $u_{\ell}$ that are equal to one is minimum. Replace $u_{\ell}$ with the
$\tilde{u}_{\ell}$ that is equal to $u_{\ell}$ except that there is 
exactly one coordinate that is equal to one in $u_{\ell}$ and is equal to zero
in $\tilde{u}_{\ell}$. Because of the minimality of $u_{\ell}$, we know that 
$\tilde{u}_{\ell}$
is different from each of $u_{1}, \ldots, u_{k}$.
The sum 
$\tilde{P}=u_{1}+ \ldots + u_{\ell-1}+\tilde{u}_{\ell}+u_{\ell+1}+ \ldots +u_{k}$
is in $S$ (also in the possible case where $\tilde{u}_{\ell}=0$).
We can now use the induction hypothesis with $\tilde{P}$ and $r-1$ and 
find the desired vector $Q$.
\end{proof}

\subsection{Proof of Theorem \ref{theorem:F} and Theorem \ref{theorem:main3}}

We start with the proof of Theorem \ref{theorem:F} which is very short.

\begin{proof}[Proof of Theorem \ref{theorem:F}.]

We are given an integer point $b=(b_{1}, \ldots, b_{d})$ 
such that $b-d^2\aov \in \Zon^d$ and 
$b_{i} \leq 2^{d-2}$ for every $1 \leq i \leq d$. We need to show that $b \in S$.

From the solution of the linear program defining the degree-sequence polytope $\Zon^d$ via its facet constraints, we conclude from $b-d^2\aov \in \Zon^d$ that $b-d^2\aov=P+W$, where $P \in S$ and 
$W=(w_{1}, \ldots, w_{d})$ is an integer vector that is a convex combination
of $d$ vectors in $C_{d}$, none of which is used in the representation of $P$
as a vector in $S$. 

In particular, we conclude that $w_{i} \leq  d$ for every $i$.
We start with $P \in S$ and apply Claim \ref{claim:1} $w_{1}$-times with $i=1$ to 
conclude that $P+w_{1}e_{1}+w_{1}\aov \in S$. We continue from there and apply Claim \ref{claim:1}
$w_{2}$-times with $i=2$
and so on until $w_{d}$-times with $i=d$ to conclude that 
$P+\sum_{i=1}^{d}w_{i}e_{i}+(\sum_{i=1}^{d}w_{i})\aov \in S$. Notice that 
$\sum_{i=1}^{d}w_{i}e_{i}=W$. Hence, $P+W+(\sum_{i=1}^{d}w_{i})\aov \in S$.
Now we apply Claim \ref{claim:2}
$d^2-(\sum_{i=1}^{d}w_{i})$ many times to conclude that 
$P+W+d^2\aov \in S$. Therefore, $b=P+W+d^2\aov \in S$.
\end{proof}

Having proved Theorem \ref{theorem:F}, we now continue with the proof of Theorem \ref{theorem:main3}.

\begin{proof}[Proof of Theorem \ref{theorem:main3}.]
From the solution of the linear program defining the degree-sequence polytope $\Zon^d$ via its facet constraints, and 
because we assume that $b-(d+1)^2(\sum_{i \in I}e_{i}-\sum_{j \in J}e_{j})) \in 
\Zon^d$, we conclude that 
$b-(d+1)^2(\sum_{i \in I}e_{j}-\sum_{j \in J}e_{j}))=P+W$, where 
$P \in S$ and 
$W=(w_{1}, \ldots, w_{d})$ is a vector in $\zplus$ 
that is a fractional convex combination
of $d$ vectors in $C_{d}$. In particular $0 \leq w_{i} \leq d-1$ for every
$1 \leq i \leq d$.

Let $r=\sum_{i=1}^{d}b_{i}\, (\text{mod } d+1)$.
By Claim \ref{claim:modulu}, there are $Q=(q_{1}, \ldots , q_{d}) \in S$ 
and $R=(r_{1}, \ldots, r_{d})$ such 
that $P=Q+R$, $\sum_{i=1}^{d}b_{i} \equiv \sum_{i=1}^{d}q_{i}\, (\text{mod } d+1)$, and
$\sum_{i=1}^{d}r_{i}=r \leq d$.

Hence 
\begin{equation}\label{eq:Q}
b-(d+1)^2\left(\sum_{i \in I}e_{i}-\sum_{j \in J}e_{j}\right)=P+W=Q+(R+W).
\end{equation}

Equation (\ref{eq:Q}) together with $\sum_{i=1}^{d}b_{i} \equiv \sum_{i=1}^{d}q_{i}\, (\text{mod } d+1)$ 
imply now that
$\sum_{i=1}^{d}(r_{i}+w_{i}) \equiv 0\, (\text{mod } d+1)$.
Write $\sum_{i=1}^{d}r_{i}+w_{i}=(d+1)a$ for an appropriate integer $a$ 
and notice that $0 \leq a \leq d+1$.

Considering (\ref{eq:Q}) again, we see that 
\begin{equation}\label{eq:diff}
\sum_{i=1}^{d}b_{i}-\sum_{i=1}^{d}q_{i}=a(d+1)+(d+1)^2(|I|-|J|)=
(d+1)(a+(d+1)(|I|-|J|)).
\end{equation}

We will apply Claim \ref{claim:1} successively on $Q$ to obtain new
points in $S$, the sum of whose coordinates increases by $d+1$ at each 
application of Claim \ref{claim:1}. 
If we can make $a+(d+1)(|I|-|J|)$ successive applications of 
Claim \ref{claim:1}, then
we result with a vector 
$Q^{*}=(q^{*}_{1}, \ldots, q^{*}_{d}) \in S$ such that 
$\sum_{i=1}^{d}b_{i}=\sum_{i=1}^{d}q^{*}_{i}$.
We will show that we can indeed apply Claim \ref{claim:1} 
$a+(d+1)(|I|-|J|)$ times.

Recall that when we apply
Claim \ref{claim:1} for a point in $Y \in S$ and an index $i$ for which 
the $i$-th coordinate of $Y$ is smaller than $2^{d-2}$, then 
we conclude that $Y+\aov+e_{i} \in S$. Hence, we moved from $Y$ to another
point in $S$ having each coordinate of $Y$ increased by one except
the $i$-th coordinate that is increased by two. 
It is extremely important to emphasize that we apply 
Claim \ref{claim:1} successively $a+(d+1)(|I|-|J|)$ times,
but we do it in such a way that every index from $I$ is chosen roughly the
same number of times and specifically at most 
$\frac{a+(d+1)(|I|-|J|)}{|I|}+1$ many times. We only need to show
that as long as we keep this rule, the $i$-th coordinate of the resulting
point in $S$ at every step remains smaller than $2^{d-2}$ for every 
$i \in I$. This is important in order to 
guarantee that we can continue and apply Claim \ref{claim:1} the number 
of times we need.

Consider $j \in J$. After every application 
of Claim \ref{claim:1}, every time with an index $i \in I$,
the $j$'th coordinate of the point in $S$ that we obtain 
increases by $1$. 
Because of (\ref{eq:Q}) we conclude that 
\begin{align*}
& b_{j}-q^{*}_{j} = -(d+1)^2+r_{j}+w_{j}-(a+(d+1)(|I|-|J|)) \\
&\leq -(d+1)^2+d+(d-1) 
\leq 0. 
\end{align*}

Now consider $i \in I$. 
The analysis in this case is a bit more involved.
After every application 
of Claim \ref{claim:1}, every time with an index $i$ in $I$,
each of the coordinates of the points in $S$ that we obtain 
increases by one except for the $i$-th coordinate that increases by two.
Because we do not apply Claim \ref{claim:1} with the same index more than 
$\frac{a+(d+1)(|I|-|J|)}{|I|}+1$ times, we conclude from 
(\ref{eq:Q}) the following inequalities.
\begin{eqnarray}
b_{i}-q^{*}_{i} \\  \geq & (d+1)^2+r_{i}+w_{i}-(a+(d+1)(|I|-|J|))
-\frac{a+(d+1)(|I|-|J|)}{|I|}-1\nonumber\\
\geq& (d+1)^2-(a+(d+1)(|I|-|J|))(1+\frac{1}{|I|})-1\nonumber\\
=& (d+1)^2-(a+(d+1)(2|I|-d))(1+\frac{1}{|I|})-1\nonumber\\
 \geq & (d+1)^2-(d+1)(2|I|-d+1)(1+\frac{1}{|I|})-1.\label{eq:f1}
\end{eqnarray}

The right hand side of (\ref{eq:f1}) is minimized when $|I|$ is maximum.
Because we assume $|J|>0$, we have $|I| \leq d-1$.
Therefore,
\begin{align*}
b_{i}-q^{*}_{i} \geq (d+1)^2-(d+1)(d-1)(1+\frac{1}{d-1})-1= d > 0.  
\end{align*}

In particular, for every index $i \in I$ the value of the 
$i$-th coordinate of the point in $S$, which we get after each of the (at most)
$\frac{a+(d+1)(|I|-|J|)}{|I|}+1$
applications of 
Claim \ref{claim:1}, is strictly smaller than $b_{i}$ and consequently
strictly smaller than $2^{d-2}$. This shows that
Claim \ref{claim:1} can indeed be applied again and again, 
$a+(d+1)(|I|-|J|)$ times, until getting the 
point $Q^*$ in $S$.

We claim that we can use Lemma \ref{lemma:simple} with $Q^*$ in the role of $Q$
and with the partitioning of the indices into $I$ and $J$ to
conclude that $b \in S$.
We need to show that the three conditions in Lemma \ref{lemma:simple}
are satisfied. We have already seen that $q^{*}_{i} \leq b_{i}$ for every
$i \in I$ and that $q^{*}_{j} \geq b_{j}$ for every $j \in J$. We know also that $\sum_{i=1}^{d}b_{i}=\sum_{i=1}^{d}q^{*}_{i}$
These are exactly the last two conditions in Lemma \ref{lemma:simple}.
To see that the first condition is satisfied, we recall that 
for every $i \in I$ we have $b_{i} \leq 2^{d-2}$
and for every $j \in J$ we have $b_{j} \geq 2^{d-2}$. Hence,
for every $i \in I$ and $j \in J$ we have $b_{i} \leq b_{j}$.
We can therefore conclude now from Lemma \ref{lemma:simple} that $b \in S$.
 \end{proof}

\section{Integer points in degree sequence-polytope that are not degree sequences} \label{sec:non-realizable}

The following theorem provides a family of  integer points of the degree-sequence polytope that are not degree-sequences, i.e. integer lattice points that cannot be written as a sum of distinct vectors of $C_d$. 
As we shall see, these non-realizable points constructed in the proof of Theorem \ref{theorem:main}, lie on the boundary of the degree-sequence polytope. The specific construction in the proof of Theorem \ref{theorem:main} gives rise to 
more unrealizable points even in the same dimension $d=\binom{8}{2}+\binom{8}{4}=98$. 
We can change the order of columns in the matrix $B$, for example. 
Furthermore, it can be easily seen how the construction in the proof of Theorem \ref{theorem:main} 
extends to other families of non-realizable points by taking other combinations of binomial 
coefficients in higher dimensions. One can also consider the possibilities of lifting lower dimensional 
non-realizable points to higher dimensional ones.

\begin{theorem}\label{theorem:main}
For $d \geq \binom{8}{2}+\binom{8}{4}$ the zonotope $\Zon^d$ contains 
integer points that are not 
realizable.
\end{theorem}

\begin{proof}
We prove the theorem for $d=\binom{8}{2}+\binom{8}{4}$. 
This will immediately imply the claim for any greater dimension, as $\Zon^d$ can be 
embedded in $\Zon^{d'}$ for $d'>d$ by adding $d'-d$ zero coordinates to every 
point. Hence, for the rest of the proof we fix $d=\binom{8}{2}+\binom{8}{4}$.

Recall that for every $k$ we let $C_k=\{0,1\}^k$ denote the set of of all vertices of the unit cube of dimension $k$. 

We construct eight vectors $v_{1}, \ldots, v_{8}$ 
in $C_{d}$ in the following way.
Let $B$ be the $8 \times (\binom{8}{2}+\binom{8}{4})$ matrix whose first $\binom{8}{2}$ columns are all the vectors in $C_{8}$ that contain precisely two coordinates
that are equal to one. Then the next $\binom{8}{4}$ columns of $B$
are all the vectors in $C_{8}$ that contain precisely four coordinates
that are equal to one.

We take $v_{1}, \ldots, v_{8}$ to be the eight rows of the matrix $B$.
Notice that indeed, $v_{1}, \ldots, v_{8} \in C_{d}$. Let $U_{0}$ denote all the vectors in $C_{d}$ that are in the linear span of 
$v_{1}, \ldots, v_{8}$. We will later show that $U_{0}$ consists only of 
$v_{1}, \ldots, v_{8}$, apart from the zero vector, of course. 

Let $c \in \mathbb{R}^d$ be a vector that is perpendicular to the linear span
of $U_{0}$, but to no vector in $C_{d} \setminus U_{0}$. Such a vector $c$ 
must exist. This is because the space of all vectors perpendicular to the linear
span of $U_{0}$ has dimension $d-k$, where $k$ is the dimension of the linear
span of $U_{0}$. The dimension of the space of all vectors perpendicular
to the linear span of $U_{0}$ and any vector not in the linear span of $U_{0}$
has dimension $d-k-1$. A space of dimension $d-k$ cannot be covered by 
a finite union of spaces of dimension $d-k-1$.

Define $U_{1} \coloneqq \{v \in C_{d} \mid \langle v,c \rangle >0\}$ and 
$U_{-1} \coloneqq \{v \in C_{d} \mid \langle v,c \rangle <0\}$.
Notice that $C_{d}$ is the disjoint union of $U_{-1}$, $U_{0}$, and $U_{-1}$.

Let $z_{0} \coloneqq \sum_{v \in U_{1}}v$. Let $b$ be the vector
$b \coloneqq \frac{1}{2}(v_{1}+v_{2}+\ldots+v_{8})$. Notice that $b$ is the vector
whose first $\binom{8}{2}$ coordinates are equal to $1$ and its next 
$\binom{8}{4}$ coordinates are equal to $2$.

Clearly, $z_{0}+b $ is contained in the degree-sequence polytope $\Zon^d$. We claim that $z_{0}+b$ is not realizable.
That is, we claim that $z_{0}+b$ cannot be written as a sum of distinct vectors
from $C_{d}$.

Assume to the contrary that $z_{0}+b$ is realizable and and let $U_{b} \subset C_{d}$ be a set such that $z_{0}+b=\sum_{v \in U_{b}}v$.
Notice that $ \langle z_{0}+b, c \rangle = \langle z_{0},c \rangle$. Observe that the hyperplane through $z_{0}$
that is perpendicular to $c$ is a supporting hyperplane of the degree-sequence polytope $\Zon_{d}$ and
$z_{0}+b$ is a point in this supporting hyperplane.
Hence, it must be that every vector
in $U_{1}$ is in $U_{b}$ and no vector in $U_{-1}$ is in $U_{b}$.
This implies that $b$ is the sum of vectors in $U_{0}$.

\begin{lemma}\label{claim:m}
$U_{0}$ consists only of $v_{1}, \ldots, v_{8}$ and the zero vector.
\end{lemma}

\begin{proof}
Recall that $U_{0}$ is the set of all vectors in $C_{d}$ that belong to the 
linear span of $v_{1}, \ldots, v_{8}$.
Consider any linear combination 
\begin{equation}\label{eq:l}
\lambda_{1}v_{1}+ \ldots +\lambda_{8}v_{8} \in C_{d}
\end{equation}
of $v_{1}, \ldots, v_{8}$ and assume it belongs to $C_{d}$.
Then, by considering the first $\binom{8}{2}$ coordinates of (\ref{eq:l}),
we conclude that for every $1 \leq i < j \leq 8$ we have 
$\lambda_{i}+\lambda_{j} \in \{0,1\}$.
By considering the next $\binom{8}{4}$ coordinates of (\ref{eq:l}),
we conclude that for every $1 \leq i<j<k<\ell \leq 8$ we have
$\lambda_{i}+\lambda_{j}+\lambda_{k}+\lambda_{\ell} \in \{0,1\}$.

Assume first that there are no $1 \leq i < j \leq 8$ such that 
$\lambda_{i}+\lambda_{j}=1$. Then for every $1 \leq i < j \leq 8$
we have $\lambda_{i}+\lambda_{j}=0$ and consequently $\lambda_{i}=0$ for
every $i$.

Therefore, we may assume without loss of generality that 
$\lambda_{1}+\lambda_{2}=1$. For every $3 \leq i < j \leq 8$
we have $\lambda_{1}+\lambda_{2}+\lambda_{i}+\lambda_{j} \in \{0,1\}$.
Consequently, $\lambda_{i}+\lambda_{j}$ must be equal to zero for every 
$3 \leq i < j \leq 8$. This implies $\lambda_{i}=0$ for $3 \leq i \leq 8$.
Observe now that $\lambda_{1}=\lambda_{1}+\lambda_{3} \in \{0,1\}$
and similarly $\lambda_{2} \in \{0,1\}$. This, together with $\lambda_{1}+\lambda_{2}=1$, implies 
that the linear combination in (\ref{eq:l}) is either equal to $v_{1}$ or to 
$v_{2}$.
\end{proof}

We know that $b$ must be the sum of vectors in $U_{0}$. By Lemma \ref{claim:m}, $U_{0}$
consists only of $v_{1}, \ldots, v_{8}$ and the zero vector. 
Therefore, in order to reach a contradiction, we show that $b$ cannot be written as a sum of 
vectors from $v_{1}, \ldots, v_{8}$. This is easy to see by inspection, or
by observing that $b$ is not any one of $v_{1}, \ldots, v_{8}$.
Furthermore, any sum of two (or more) of $v_{1}, \ldots, v_{8}$ will necessarily 
have one of the first $\binom{8}{2}$ coordinates being greater than 
or equal to $2$, while the first $\binom{8}{2}$ coordinates of $b$ are all
equal to one. 

This concludes the proof of Theorem \ref{theorem:main}, showing that $z_{0}+b$ is an integer vector in $\Zon^d$
that is not realizable.
\end{proof}

The construction presented in this section shows that there are integer lattice points in the degree sequence polytope for $d \geq 98$ that are not realizable. In contrast to this result, computational experiments showed that there are no 
non-realizable points in the degree-sequence polytope for $d \leq 9$.
This leads to the unanswered question whether the 
construction we found is minimal in the dimension. 

 \section{Complexity of \DSP and the linear optimization problem over degree sequences} 
 \label{sec:compl-dsp-line}

 The $3$-uniform hypergraphic degree sequence problem is the problem of deciding whether  $b \in \mathbb{Z}^{d}_{\geq 0}$ is a degree sequence of a $3$-uniform hypergraph.  This problem is known to be $\mathrm{NP}$-complete~\cite{MR3840885}. 
 We now show how to  transform $b$ into an instance of \DSP, showing that \DSP is $\mathrm{NP}$-hard.

 Consider the vector $w \coloneqq (-3,1,1, \dots ,1)^{\top} \in \mathbb{R}^{d+1}$, and let $A_{+} \coloneqq \{v \in C_{d+1} : \langle v,w \rangle > 0 \}$. We define
 $A_{0} \coloneqq \{v \in C_{d+1} : \langle v,w \rangle = 0 \}$.
The set $A_{0}$ consists  precisely those vectors in $C_{d+1}$ whose first coordinate is equal to $1$ and precisely three other coordinates are equal to $1$, that is $A_{0}= \{v \in C_{d+1} : v = (1, x^{\top})^{\top}, ~~\lVert x \rVert_1=3 \}$. Define $z \coloneqq \sum_{v \in A_{+}} v$. Then $w$ is perpendicular to the affine hyperplane $H \coloneqq z + \{v \in \Zon^{d+1} : \langle v,w \rangle = 0 \} $. Notice that $A_{0} \subseteq H$. 

 \begin{lemma}\label{lemma:red}
 A sequence $b \in \mathbb{N}^d$ is a degree sequence of a $3$-uniform hypergraph if and only if the integer point  $z + (\frac{\lVert b \rVert_1}{3}, b^{\top})^{\top}$ is a degree sequence. 
 \end{lemma}

 \begin{proof}
 If $b$ is a degree sequence of a $3$-uniform hypergraph, then 
 we can write $b$ as a sum of pairwise distinct vectors $b=x_{1}+ \ldots +x_{k}$ where every $x_{i} \in C_{d}$ is a $0/1$-vectors with exactly three $1$-entries and hence 
 $(1,x_{i}^{\top})^{\top} \in A_{0}$.

We have $(\frac{\lVert b \rVert_1}{3}, b^{\top})^{\top}=
\sum_{i=1}^{k}(1,x_{i}^{\top})^{\top}$. Consequently,
$z + (\frac{\lVert b \rVert_1}{3}, b^{\top})^{\top}$ is realizable in $\Zon^{d+1}$ as it is equal to
$\sum_{v \in A_{+}}v + \sum_{i=1}^{k}(1,x_{i}^{\top})^{\top}$.

 Conversely, if $z + (\frac{\lVert b \rVert_1}{3}, b^{\top})^{\top}$ is realizable, then it is equal to the sum 
 of all vectors in a subset $S \subset C_{d+1}$.  
The sum $\sum_{v \in S}v$ should belong to the supporting hyperplane 
$H$. Therefore $S$ must be equal to the union of $A_{+}$ and a subset of $A_{0}$. Because $z=\sum_{v \in A_{+}}v$, we conclude that $(\frac{\lVert b \rVert_1}{3}, b^{\top})^{\top}$ is equal 
to the sum of pairwise distinct vectors in $A_{0}$.
In particular, $b$ is equal to the sum of vectors in $C_{d}$
each of which has precisely three coordinates that are equal to $1$. Hence, $b$ is a degree sequence of a $3$-uniform hypergraph.
 \end{proof}

We observe that $z$ can be constructed in polynomial time. 
 Recall that $z$ is the sum of all vectors in $A_{+}$. The set $A_{+}$ is the set of all vectors in $C_{d+1}$ such that $\langle v,w \rangle > 0$, where 
$w \coloneqq (-3,1,1, \dots ,1)^{\top} \in \mathbb{R}^{d+1}$.

The scalar product of $w$ and a vector in $C_{d}$ is equal to 
an integer that is greater than or equal to $-3$.
For $k=-3,-2,-1,0$ we define the sets
$A_{k}=\{v \in C_{d+1} : \langle v,w \rangle=k\}$.
Then 
$$
z=\sum_{v \in C_{d+1}}v - \sum_{v \in A_{-3}}v - 
\sum_{v \in A_{-2}}v - \sum_{v \in A_{-1}}v - 
\sum_{v \in A_{0}}v.
$$

We can now easily find $z$ using the following 
equalities:
\begin{eqnarray}
\sum_{v \in C_{d+1}}v &=& (2^{d}, \ldots, 2^{d}),\nonumber\\
\sum_{v \in A_{-3}}v&=&(1,0, \ldots, 0),\nonumber\\
\sum_{v \in A_{-2}}v&=&(d,1, \ldots, 1),\nonumber\\
\sum_{v \in A_{-1}}v&=&\biggl(\binom{d}{2}, d-1,\ldots, d-1\biggl),\nonumber\\
\sum_{v \in A_{0}}v&=&\biggl(\binom{d}{3}, \binom{d-1}{2}, \ldots,
\binom{d-1}{2}\biggl).\nonumber
\end{eqnarray}
\noindent
We conclude with the following theorem.
\begin{theorem}
  \label{thr:1}
  The degree-sequence problem \DSP is $\mathrm{NP}$-hard. 
\end{theorem}

Deza et al.~\cite{MR3840885} showed that one can maximize a linear function over the set of degree sequences of hypergraphs with at most $m$ edges in polynomial-time in $d$ and  $m$. Our final result shows that this parameterization in $m$ is necessary. 

\begin{proposition}\label{prop:optimizing}
    Optimizing over the degree-sequence polytope $\Zon^d$ is $\mathrm{NP}$-hard.
\end{proposition}
\begin{proof}
  We reduce from  \emph{counting knapsack solutions}. This problem is known to be $\#P$-hard~\cite{valiant1979complexity} and defined as follows. Given $a \in \mathbb{Z}^d_{\geq 0}$, $\beta \in \mathbb{Z}_{\geq 0}$, determine
  \begin{displaymath}    
  N = \left| \{ x \in \{0,1 \}^d : a^T  x <  \beta \} \right|.\footnote{Strict inequality $<$ is justified by multiplying $a$ and $\beta$ by two and then by augmenting $\beta$ by one.}
\end{displaymath}
  Notice that for $ x \in \{0,1 \}^d$ one has  $a^T x <  \beta$ if and only if  $a^T x <  \beta - 1/2$, since $a$ and $\beta$ are integral. 

  The linear optimization problem over $\Zon^d$ is solved by summing the $0/1$ vectors with positive objective value. In other words, for 
   $w  \in  \mathbb{R}^{d}$, one has
  \begin{displaymath}
    \max\{w^T x : x \in  \Zon^{d}\} =  \sum_{ \substack{ z  \in  \{0,1\}^d  \\  w^Tz >0}} w^Tz .  
  \end{displaymath}
  To this end, let $w_1^T  =  (-c^T,  \beta)$ and $w_2^T  =  (-c^T,  \beta -1/2)$. Clearly, $x_0  \in \{0,1\}^d$ is a knapsack solution if and only if
  $w_1^T \left(
\begin{smallmatrix}
  x_0 \\1
\end{smallmatrix}\right) >0$ and this is the case if and only if $w_2^T \left(
\begin{smallmatrix}
  x_0 \\1
\end{smallmatrix}\right) >0$. If we denote the optimum values of the linear optimization problems over $\Zon^{d+1}$  with objective vector $w_1$ and $w_2$ by $OPT_1$ and $OPT_2$ respectively, then $OPT_1 = OPT_2 + N/2$. This shows that the linear  optimization problem over the degree-sequence polytope is $\mathrm{NP}$-hard. 
\end{proof}

\bibliographystyle{plainurl}
\bibliography{ms.bib}

\end{document}